\documentclass[runningheads]{llncs}
\usepackage[T1]{fontenc}
\usepackage{graphicx}
\usepackage{algorithm}
\usepackage{algorithmicx}
\usepackage{algpseudocode}
\usepackage{appendix}
\usepackage{amsmath}
\usepackage{amsfonts}
\usepackage{array}
\usepackage{balance}
\usepackage{float}
\usepackage{listings}
\usepackage{multicol}
\usepackage{multirow}
\usepackage{stmaryrd}
\usepackage{subfig}
\usepackage{tikz}
\usepackage{xcolor}
\usepackage[misc,geometry]{ifsym}
\usetikzlibrary{shapes, arrows}
\usepackage{hyperref}
\usepackage{tabularx} 
\usepackage{ragged2e} 
\usepackage{booktabs} 

\definecolor{mygreen}{rgb}{0,0.6,0}
\definecolor{mygray}{rgb}{0.5,0.5,0.5}
\definecolor{mymauve}{rgb}{0.58,0,0.82}

\begin{document}
\title{AC\texorpdfstring{$^4$}{4}: Algebraic Computation Checker for Circuit Constraints in Zero-Knowledge Proofs}
%
%

\author{
Qizhe Yang\inst{2}
\and
Boxuan Liang\inst{1}
\and
Hao Chen\inst{1}
\and
Guoqiang Li\inst{1}\thanks{Corresponding author.}
}
\authorrunning{Q. Yang et al.}
\titlerunning{AC\texorpdfstring{$^4$}{4}: Algebraic Computation Checker for Circuit Constraints in ZKPs}

\institute{
Shanghai Jiao Tong University, Shanghai, 200240, China\\
\email{\{dennis.Chen,  628628\}@sjtu.edu.cn} \and
Shanghai Normal University, Shanghai, 200234, China\\
\email{qzyang@shnu.edu.cn}
}
%
%
\maketitle              
\begin{abstract}
 \emph{Zero-knowledge proof (ZKP)} systems have surged attention and held a fundamental role in contemporary cryptography. \emph{Zero-knowledge succinct non-interactive argument of knowledge (zk-SNARK)} protocols dominate the ZKP usage, implemented through arithmetic circuit programming paradigm. However, underconstrained or overconstrained circuits may lead to bugs. The former refers to circuits that lack the necessary constraints, resulting in unexpected solutions and causing the verifier to accept a bogus witness, and the latter refers to circuits that are constrained excessively, resulting in lacking necessary solutions and causing the verifier to accept no witness. This paper introduces a novel approach for pinpointing two distinct types of bugs in ZKP circuits. The method involves encoding the arithmetic circuit constraints to polynomial equation systems and solving them over finite fields by the \emph{computer algebra system}. The classification of verification results is refined, greatly enhancing the expressive power of the system. A tool, AC\texorpdfstring{$^4$}{4}, is proposed to represent the implementation of the method. Experiments show that AC\textsuperscript{4} demonstrates a increase in the solved rate, showing a 29\% improvement over Picus and CIVER, and a slight improvement over halo2-analyzer, a checker for halo2 circuits. Within a solvable range, the checking time has also exhibited noticeable improvement, demonstrating a magnitude increase compared to previous efforts.
\end{abstract}
\section{Introduction}
\emph{Zero-knowledge proof (ZKP)} systems have long occupied a fundamental position in contemporary cryptography, ever since their inception~\cite{abc10}. They are significantly utilized in security-sensitive applications, such as smart contracts~\cite{abc98}, blockchain confidential transactions~\cite{abc73}, and digital currency~\cite{zcash}, given their capacity to bolster security and privacy. ZKP systems' potency lies in their ability to prove anything that an interactive proof system can, but without revealing any knowledge~\cite{10.1007/3-540-48184-2_4,10.1007/0-387-34799-2_4}. Therefore, these systems have assumed the role of a potent tool enabling trust and authentication while safeguarding sensitive information.

Developers frequently choose the \emph{zero-knowledge succinct non-interactive argument of knowledge (zk-SNARK)}~\cite{184425} protocol for its efficiency in setting up and proving within a zero-knowledge proof (ZKP) system. Additionally, they adopt arithmetic circuits~\cite{algebraicComplexityTheory} as the programming paradigm due to their simplicity in representing computational problems. Both Circom and halo2 are built upon this protocol and programming paradigm. To illustrate this, consider the process of declaring a boolean value in circuit programs. In high-level languages, a boolean value \(b\) can be declared directly. However, in circuit languages, a constraint must be explicitly defined to limit its possible values. For instance, the constraint \(b \times (b - 1) = 0\) ensures that \(b\) can only be 0 or 1. This transformation cannot be automatically performed by existing compilers or libraries, necessitating that developers express computations using polynomial equations.

The limitation of polynomial constraint systems gives rise to two problems. The first problem is the \emph{underconstrained problem}~\cite{10.1145/3591282}, which occurs when multiple outputs can fulfill the same system of equations with identical input values. This situation arises when the system of polynomial equations does not yield a mathematical function on the finite field, rendering the circuit underconstrained. As a result, malicious users could generate fake witnesses, deceive the verifier, and potentially steal the cryptocurrency~\cite{DisclosureRecentVulnerabilities,cashTornadoCashGot2019}. The second one is the \emph{overconstrained problem}~\cite{10.1145/3591282}, which occurs when no outputs can satisfy the system of polynomial equations or the constraint system is inconsistent for a given input. This issue can result in a non-functional circuit. Consequently, this may lead to overconstrained bugs that cause ZK-verifier servers to reject numerous legitimate requests. Such rejections could potentially trigger Distributed Denial of Service (DDoS) problems. Furthermore, it is evident that underconstrained circuits can produce more severe consequences than their overconstrained counterparts.

To this end, we conducted an investigation into the constraints on Circom or halo2 circuit programs. By reconsidering the formalization paradigm, we modeled underconstrained and overconstrained problems as algebraic inquiries. Specifically, we examined whether a system of polynomial equations over a finite field has multiple solutions. To support our motivation, we analyzed the official library of circom circuit templates provided by iden3 and collected halo2 circuit templates from GitHub Projects. Our analysis revealed that nearly 70\% of these templates are linear, and efficient algorithms exist for solving linear systems of equations. Consequently, we categorized constraint problems into diverse types based on their maximal degree. We then employed a unified arithmetic solution framework to address each type separately.

To summarize, contributions can be elaborated as follows:
    \begin{itemize}
        \item We proposed a methodology that utilizes computer algebra systems to model circuit programs as polynomial systems over finite fields, with the aim of verifying their correctness.
        \item We developed \textbf{AC\texorpdfstring{$^4$}{4}}, a highly efficient and practical tool that leverages the aforementioned framework to effectively handle larger circuit programs. 
        
        \item We constructed datasets specifically tailored for Circom and halo2 circuits. By comparing the solved rate and checking speed of AC\texorpdfstring{$^4$}{4} against other tools, we observed substantial improvements in both metrics. 
    \end{itemize}

\section{Background}

\smallskip \noindent \textbf{Zk-SNARKs.}
Non-interactive ZKP protocols, such as zk-SNARKs~\cite{6956581}, allow verifiers to be assured of a statement's truth without gaining any knowledge beyond its validity~\cite{abc10}. Blum et al. demonstrated that computational zero-knowledge can be achieved without interaction by sharing a common random string between the prover and the verifier~\cite{10.1145/62212.62222,560484}. The non-interactive protocol conveys the proof through a message from the prover to the verifier, ensuring efficiency. zk-SNARKs reign supreme in ZKP systems due to lightning-fast verification of even large programs, using tiny proofs (mere hundreds of bytes) within milliseconds.

\smallskip \noindent \textbf{Arithmetic Circuits.}
Arithmetic circuits, fundamental models for polynomial computation in complexity theory, process variables/numbers using addition and multiplication. Conceptually, arithmetic circuits offer a formal method for understanding the complexity of computing polynomials. Note that all variables mentioned in arithmetic circuits represent signals in the Circom language.

\begin{definition}[Arithmetic circuit~\cite{pmlr-v49-volkovich16}] \label{def:arithmetic_circuit_def}
An arithmetic circuit $C(\mathcal{X}) $ in the variables $\mathcal{X}=\{X_1, \ldots, X_k\}$ over the field \(\mathbb{F}\) is a labelled directed acyclic graph. 
The inputs are assigned by variables from $X$ or by constants. The internal nodes act as computational operations that add or multiply to calculate the sum or product of the polynomials on the tails of incoming edges. The output $X_{out} \in \mathcal{X}$ of a circuit is the polynomial computed at the output node. 
\end{definition}

\smallskip \noindent \textbf{Circuit Languages.} 
The growing field of ZKPs has led to the development of specialized circuit languages designed to create and manage the underlying circuits needed for these proofs. Two notable examples are the \textit{Circom}~\cite{Circom} and \textit{halo2}~\cite{opensourcebook:halo2book} frameworks.

{Circom} is a domain-specific language that leverages SNARK technology and employs arithmetic circuits as its programming paradigm. Programmers define circuits using specific operators: \(===\) for equality checks, \(<--\) or \(-->\) for signal assignment, and \(<==\) or \(==>\) for simultaneous assignment. These circuits conform to the \emph{rank-1 constraint system (R1CS)} format. The R1CS format is expressed as \(A \times B + C = 0\), where \(A\), \(B\), and \(C\) are linear combinations of signals. After compiling the circuit into R1CS format and computing a witness using the input, the \textit{snarkjs} tool can be employed to generate and verify a ZKP for the specified input. 

{Halo2} provides a more flexible and scalable method for constructing ZKPs. It employs PLONKish arithmetization~\cite{plonk}, which supports recursive proof composition, thereby enhancing scalability. In \textit{halo2}, circuits are built using polynomials and lookups. This construction allows for more efficient proof generation and verification.

It is reasonable to conclude that arithmetic circuits in both circom and halo2 can be treated as sets of polynomial equations, despite differing in their representations, methods of construction, and evaluation. In particular, halo2 supports constraints that are typically quadratic or higher. The specific implementation and optimization goals determine the exact nature of these constraints.
We address circuit constraint verification by extending arithmetic circuit definitions with constraints.

\begin{definition}[Constraint in Circom]
A constraint \(e\) is a quadratic equation over a finite field \(F_p\):
    \[\mathcal{A}\mathcal{X} \times \mathcal{B}\mathcal{X} + \mathcal{C}\mathcal{X} = 0\]
    where \(\mathcal{A}, \mathcal{B}, \mathcal{C}\) are coefficient vectors over \(F_p\), \(\mathcal{X}\) is a vector of variables over \(F_p\), and the cross product \(\times\) means the product of 2 scalars \(\pmod p\).
\end{definition}

A halo2 circuit constraint is more general.
\begin{definition}[Constraint in halo2]
A constraint in halo2 is a polynomial equation over a finite field \(F_p\):
\[
\sum_{i=0}^{n} \sum_{j=0}^{n} a_{i,j} \cdot x_i \cdot x_j + \sum_{i=0}^{n} b_i \cdot x_i + c = 0
\]
where \(n\) is number of variables, \(a_{i,j}, b_i, c \in F_p\) are the coefficient scalars, and \(x_i \in F_p\) are the variables.
\end{definition}
\begin{remark}
Since any higher-degree (\(\ge 3\)) polynomial equation can be reduced to a system of quadratic equations by introducing auxiliary variables, we restrict attention to constraints of degree at most \(2\). For example, the constraint \(x^3 + c = 0\) can be rewritten as $xy+c =0$ and $x^2-y=0$.
\end{remark}
\begin{definition}[Arithmetic circuit with constraints] \label{def:constraint_circuit_def}
An arithmetic circuit with constraint can be defined as a tuple 
\(\langle C(\mathcal{X}), E \rangle \), where \(C(\mathcal{X})\) 
 is an arithmetic circuit on variable set $\mathcal{X}$. $\mathcal{X}$ is the union of two disjoint sets \texorpdfstring{$K$}{4} and \(U\), where \texorpdfstring{$K$}{4} represents known input variables, and \(U=T\cup O\) represents unknown variables including the non-constant variables including temporal variables \(T\) and output variables \(O\). \(E\) includes all the constraints of \(C(\mathcal{X})\).
\end{definition}

\smallskip \noindent \textbf{Computer algebra.}
\emph{Computer algebra}~\cite{DBLP:conf/cascon/BrightKG19} revolves around the manipulation of algebraic expressions and various mathematical entities. This field has spurred the creation of computer algebra systems capable of autonomously tackling a broad spectrum of theoretical and practical mathematical challenges. A contemporary computer algebra system boasts a diverse array of functionalities, encompassing Gr\"{o}bner bases, cylindrical algebraic decomposition, lattice basis reduction, linear system solving, arbitrary and high-precision arithmetic, interval arithmetic, linear and nonlinear optimization, Fourier transforms, Diophantine equation solving, among others. These capabilities are indispensable for both theoretical investigations and practical applications.

\smallskip \noindent \textbf{Gr\"{o}bner basis.}
A Gr\"{o}bner basis is a collection of multivariate polynomials possessing favorable algorithmic characteristics. Any set of polynomials can be converted into a Gr\"{o}bner basis. This transformation encompasses three well-known methods: Gaussian elimination for solving systems of linear equations, the Euclidean algorithm for finding the greatest common divisor of two univariate polynomials, and the Simplex algorithm for linear programming. Algorithms on Gr\"{o}bner basis are widely used in mathematical and verification tools. The computation of a Gr\"{o}bner basis can be performed using \emph{Buchberger's algorithm}~\cite{definition:GrobnerBasis}, while \emph{Faug\`{e}re F\textsubscript{5} algorithm}~\cite{f5algorithm} is a significant improvement over the former. It introduces the concept of signatures to predict and eliminate unnecessary computations, which can potentially increase the efficiency of Gr\"{o}bner basis computation.

\section{Problem Statements}

Detecting overconstrained or underconstrained circuits in circuit languages boils down to an algebraic inquiry into the number of solutions for a polynomial equations system over a finite field. This inherent connection naturally leads us to model circuits as such systems and leverage algebraic methods for analysis. Our method follows the classification-and-discussion scheme. Our strategy categorizes circuits by degree and applies specialized solutions within each category, leading to significant efficiency and scalability gains.

\subsection{Constraint Problems}
The construction of arithmetic circuits is plagued by two critical challenges: overconstrained and underconstrained circuits. Programmer-authored circuits, even from seasoned ZKP developers, often bear inadvertently miscalibrated constraints, leading to these issues. While completely eradicating these issues remains a persistent hurdle, current approaches fall short of effectiveness. This subsection tackles this double-edged sword head-on, proposing a novel approach to ensure the construction of robust and reliable arithmetic circuits.

\smallskip \noindent \textbf{Overconstrained bugs.}
As described in the work by Labs et al.~\cite{labs2023SecurityConcernsZeroKnowledge}, overconstrained circuits are unable to uphold the completeness of ZKP. A circuit program is deemed overconstrained when its corresponding polynomial constraints are overconstrained, indicating that the circuit cannot produce any valid outputs regardless of the inputs provided. This inconsistency in the constraint system for a given input characterizes the state of being overconstrained. Despite their deceptive simplicity in concept and the misconception that they are less critical than other types of attacks, denial of service (DoS) attacks are often underappreciated regarding their security implications. Notably, DoS attacks require neither high skill nor deep knowledge,, contributing to their low cost for malicious attackers as highlighted by Zcash~\cite{ZcashDenialOfServiceAttack}.

\smallskip \noindent \textbf{Underconstrained bugs.}
Recent studies have discovered that underconstrained bugs in arithmetic circuits can lead to various security threats, including the ability for attackers to forge signatures, steal user concurrency, or create counterfeit crypto-coins. A circuit program is classified as underconstrained when the corresponding polynomial constraints are also underconstrained. This condition arises when there exists an input \(in\) for which multiple distinct outputs \(out_{1}, out_{2}, \ldots\) can be produced, and the circuits \(C(in, out_1), C(in,out_2),\ldots\) are all considered valid. In essence, an underconstrained circuit program may unintentionally permit the acceptance of unwanted signals, making it susceptible to exploitation by malicious entities.

\subsection{Categories of Elements}
In our method, categorization is essential to account for the different types of variables in arithmetic circuits. Variables can be classified based on their source: some are determined by external inputs, while others depend on internal variables. It is also important to categorize variables according to their degree. Due to the computational complexity of verifying certain circuits, we can only provide coarse-grained results. This limitation underscores the importance of categorizing the outcomes as well.

\smallskip \noindent \textbf{Category of Variables.}
The variables in a circuit program consist of the input signals \(I\), temporary signals \(T\), and output signals \(O\). The user provides the values for \(I\), which can be regarded as known values \texorpdfstring{$K$}{4}. On the other hand, the variables in \(O \cup T\), whose values are not constant, are regarded as unknown variables \(U\) within the constraints. The key issue of the constraint problem is identifying the unique characteristics of the \(O\) variables.

\smallskip \noindent \textbf{Category of Constraints and Circuits.} \label{category_of_c_c}
The constraint \(e\) is an algebraic polynomial equation whose degree is under 2 over a finite field \(F\), while the circuits are a system of algebraic polynomial equations whose degree is under 2 over a finite field, with their generator variables being \(U\).

The three types of constraints to consider are precise linear equations, \texorpdfstring{$K$}{4}-coefficient linear constraints, and higher constraints. A precise linear constraint is defined by having linear generator variables with constant coefficients. Conversely, a \texorpdfstring{$K$}{4}-coefficient linear equation includes at least one generator variable with a coefficient represented by the known variables \texorpdfstring{$K$}{4}, and none of the generator variables' coefficients are represented by the unknown variables \(U\). Lastly, a higher equation is distinguished by containing at least one generator variable with a coefficient represented by the unknown variables \(U\).

The system of constraints, namely circuits, can be categorized into three types based on the constraints outlined above: precise linear circuits, \texorpdfstring{$K$}{4}-coefficient circuits, and higher-order circuits. A precise linear circuit consists solely of precise linear constraints, while a \texorpdfstring{$K$}{4}-coefficient circuit must contain at least one \texorpdfstring{$K$}{4}-coefficient linear constraint and no higher-order constraints. Finally, a higher-order circuit should include at least one higher-order constraint.

\subsection{Problem Reduction}
This section demonstrates the reduction of the under/overconstrained problems of a circuit to the number of solutions of the corresponding polynomial equations.

\begin{definition}[Polynomials over finite fields]
    For a finite field \(\mathbb{F}\) and formal variables \(X_1, \ldots, X_k\), \(\mathbb{F}[X_1, \ldots, X_k]\) denotes the set of polynomials in \(X_1, \ldots, X_k\) with coefficients in \(\mathbb{F}\). By taking the variables to be in \(\mathbb{F}\), a polynomial \(f \in \mathbb{F}[X_1, \ldots, X_k]\) can be viewed as a function from \(\mathbb{F}^k \rightarrow \mathbb{F}\).
\end{definition}

To reduce the circuit constraint problems to the number of solutions of polynomials, 
we have the following proposition:
\begin{proposition}
[Circuit-Poly Reduction]
For each circuit $A = \langle C(X), e \rangle \in \langle C(\mathcal{X}), E \rangle$, there exists $A^{\prime} \in \mathbb{F}[X_1, \ldots, X_k]$, such that $A^{\prime} = \mathbb{M}(A)$ , where $\mathbb{M}$ is a bijection function from a circuit to polynomial equations over finite fields.
\end{proposition}

A bijection can be constructed between the formal variables of polynomials and the variables of a circuit because their domain is the same finite field \(\mathbb{
F}\). The meta operator set \(O_c\) of polynomial circuits is \(\{+, \times, \sim \}\) \footnote{\(\sim\) is the complementary operation on finite fields} and the meta operator set \(O_p\) of a polynomials are \(\{+, \times, \sim\}\), so \(O_c = O_p\). Because the additive group over finite fields is an Abelian group under `+', the operation of shifting the terms does not change the properties of the original variables and constraints. However, a constraint of a polynomial circuit can only be constant, linear, or higher, while polynomials exceed the limit of the degree of variables. Consequently, the constraints of a polynomial circuit are a subset of polynomials.

Based on the argument above, the complete checking of the constraint status of the circuits can be achieved by solving polynomials. More concretely, the constraint problem corresponds to figuring out the number of solutions for polynomials.

\section{Methodology}

\subsection{AC\texorpdfstring{$^4$}{4} Framework}

\begin{figure}[htbp]
    \centering
    \includegraphics[width = .6\linewidth]{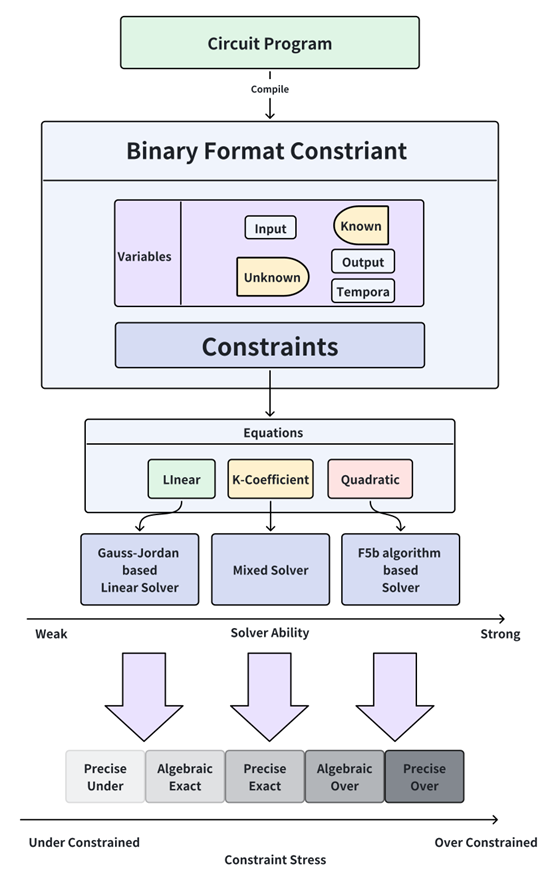}
    \caption{The framework of AC\texorpdfstring{$^4$}{4}.}
    \label{fig:arch}
\end{figure}

Fig. \ref{fig:arch} illustrates the overall workflow of our proposed tool, \(\text{AC}^4\), which receives a circuit as the input, and compiles the program in debug mode into an R1CS format file along with a sym file. To retrieve the variables and constraints from the compiled files, a task-specific parser is implemented to handle those compiled files, focusing on unknown variables within constraints. After that, the extracted constraints are then simplified into equations. A categorizing mechanism is leveraged in Section \ref{category_of_c_c} to determine the specific category of the circuit according to its simplified equation system. Based on the category result of circuits, three types of solvers is employed to address them respectively.
In the context of a precisely pure linear circuit, the system of linear equations can be represented in matrix form. While it may seem natural to rely on rank to determine the number of solutions, it is more important to prioritize verifying the uniqueness of the output variable set. Therefore, solving the system demands a suitable method for linear equations. Specifically, in \(\text{AC}^4\), the \textit{Gauss-Jordan method}\cite{siahaan2023study} is utilized to obtain the solution of the circuits as well as to ascertain the uniqueness of the set of output variables. 

The system of \texorpdfstring{$K$}{4}-coefficient linear equations corresponding to an algebraically linear circuit can be represented in matrix form. However, it is imperative to note that algebraic computation does not inherently account for the division by zero problem. Consequently, for certain specific inputs \(I\), the rank of the matrix may decrease, potentially altering the outcome of \(\text{AC}^4\). Rank reduction risks are mitigated by strategically identifying and treating special input cases. Our approach addresses this challenge through heuristic approaches to improve the efficiency and accuracy. 

For higher-order circuits, we solve the coefficients and verify if the circuit will be underconstrained with specific inputs. Then, we use \(\text{AC}^4\) to directly solve the systems of polynomial equations with Gr\"{o}ebner basis\cite{BUCHBERGER2006475} and check the uniqueness of the output variables. 

Here, an example will be used to demonstrate our workflow and enhance readers' comprehension. The template in Fig. \ref{code:decoder} refers to the Circom circuit Decoder, which converts a w-ary number into a one-hot encoded format.
\begin{figure}[!ht]
    \centering
\begin{lstlisting}
template Decoder(w) {
    signal input inp;
    signal output out[w], success;
    var lc=0;
    for (var i=0; i < w; i++) {
        out[i] <-- (inp == i) ? 1 : 0;
        out[i] * (inp-i) === 0;
        lc = lc + out[i];}
    lc ==> success; 
    success * (success -1) === 0;
}
component main = Decoder(2);
\end{lstlisting}
    \caption{Binary to One-Hot Decoder Circuit}
    \label{code:decoder}
\end{figure}
Upon setting \(w = 2\), the \textit{Decoder(2)} emerges as a 2-bit decoder. This circuit is then transformed into a format denoted as \(C(K, U, O, E\), where the sets are defined as follows: \(K = \{inp\}\), \(U = \{out_1, out_2, success\}\), \(O = \{out_1, out_2, success\}\), and \(E = \{out_0 \times inp = 0, out_1 \times (inp - 1) = 0, success \times (success - 1) = 0\}\).

Based on the classification of circuits mentioned above, it is evident that this circuit is not linear. To analyze this, we begin by treating the variable \(inp\) as a coefficient and proceed to solve for the undetermined coefficients of each equation. This process yields the solutions \([\{inp = 0\}, \{inp = 1\}]\). When substituting \(\{inp = 0\}\) into \(E\), the resulting equations are \(E = \{0 = 0, out_1 \times (inp - 1) = 0, success \times (success - 1) = 0\}\). Notably, \(out_0\) can take on any value over the finite field, indicating that the circuit is underconstrained.

\subsection{Adaptive Checking with Relaxation and Partial Results}
AC\texorpdfstring{$^4$}{4} is designed to provide precise results when analyzing circuits. However, when dealing with large or complex circuits, it might produce an ``unknown'' outcome due to the scalability. To handle these situations, AC\texorpdfstring{$^4$}{4} relaxes its strict checking conditions and instead returns partial, but still meaningful, results.

\begin{figure}[!ht]
    \centering
    \includegraphics[width = .95\linewidth]{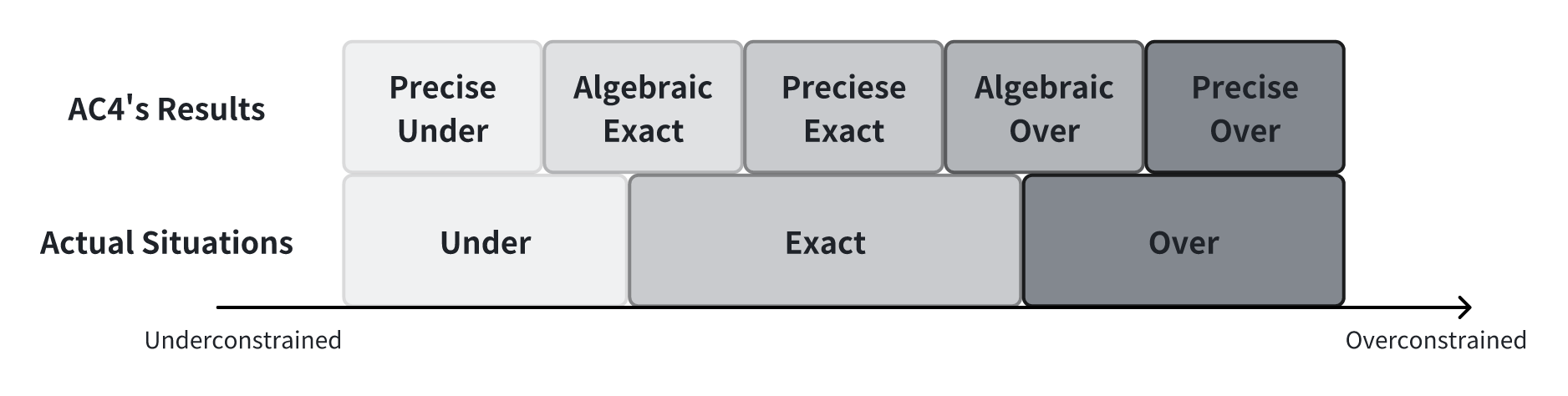}
    \caption{The category of results from AC\texorpdfstring{$^4$}{4}}
    \label{fig:stress}
\end{figure}

In certain cases, the input coefficients in circuit equations can be difficult to determine as zero or non-zero. To simplify, AC\texorpdfstring{$^4$}{4} treats all combinations of input and constant values as non-zero, allowing for a more straightforward algebraic result. This means the coefficient can be reduced in the field, thus yields a workable solution, even if not fully precise. Under this assumption, the overconstrained and exact-constrained results may not align with the actual situation of the circuits. We name the former as \emph{algebraic} results. In contrast, by \emph{precisely}, we mean that the result are directed obtained without the above assumption.
Hence, as depicted in  Fig. \ref{fig:stress}, AC\texorpdfstring{$^4$}{4} generates five practical categories of results: algebraic overconstrained, precisely overconstrained, algebraic exact-constrained, precisely exact-constrained, and precisely underconstrained. 

While its primary objective is to provide precise results during circuit analysis, AC\texorpdfstring{$^4$}{4} recognizes the challenges posed by large circuits with numerous parameters. In such scenarios, where a precise result may be unattainable, AC\texorpdfstring{$^4$}{4} adjusts its approach by relaxing the strictness of its checking conditions. This adaptability enables AC\texorpdfstring{$^4$}{4} to yield a meaningful partial result, ensuring that even with complex circuits, users can still gain valuable insights into the circuit's behavior.

To provide more detailed feedback on unsolvable circuits, AC\texorpdfstring{$^4$}{4} refines its result classification. It introduces two intermediate categories, algebraic exact-constrained and algebraic overconstrained, which offer additional insights. These categories help analysts understand whether a circuit may be underconstrained (too flexible) or overconstrained (no valid solutions) by examining the algebraic structure of the equation matrix.

The relationship between AC\texorpdfstring{$^4$}{4}’s results and the actual circuit behavior is also cross-referenced with ground truth. Three categories—precisely underconstrained, precisely exact-constrained, and precisely overconstrained-directly correspond to whether a circuit is under, exact or overconstrained. However, the algebraic categories offer more nuanced information. For example, an algebraically exact-constrained circuit might be either underconstrained or exact-constrained in reality, while algebraic overconstrained cases could actually be exact-constrained or overconstrained.

{Unknown} results may stem from circuit size, computational complexity, or the inherent NP-hardness of solving polynomial equations over finite fields.

By relaxing the strict checking conditions, AC\texorpdfstring{$^4$}{4} can filter out cases that are definitely under or overconstrained, providing more efficient and useful results even when a full solution isn't possible.

\subsection{Checking Circuits: Category-Based Algorithms}

From above, we divide the overall circuit verification problem space into three types: precisely linear, \texorpdfstring{$k$}{4}-coefficient linear and higher-order separately. In this section, we introduce those algorithms in detail.

\smallskip
\noindent \textbf{Precisely linear.}
The majority of circuits are precisely linear, so they can be represented by a coefficient matrix \(A\) and a constant vector \(b\). While it is commonly thought to use the rank of the equation matrix to determine the number of solutions, the solver's objective is only to verify the uniqueness of output variables. In contrast, rank-reducing methods impose restrictions on the uniqueness of all variables, including temporary variables. Using the Gauss-Jordan method to get the rank of the corresponding matrix is analogous to getting the solution of the system of equations. Consequently, there is no significant difference in the time complexity. The pesudo-code is given in Algorithm \ref{algo:linear_check}.

\begin{algorithm}[!ht]
    \caption{Algorithm for checking linear circuits} \label{algo:linear_check}
    \begin{algorithmic}[1]
    \Procedure{\textsc{CheckLinearCircuit}}{$\langle C(\mathcal{X}), E \rangle$}
        \State \(A, b \gets \textsc{LinearEquationsToMatrix}(E, U)\) \Comment{\(A\) is the coefficient matrix and \(b\) is the constant matrix}
        \State \(sols \gets \textsc{GaussJordanSolve(A, b)}\)
        \If{\textsc{IsEmpty}(sols)}
            \State \textbf{return} \textit{precisely overconstrained}
        \ElsIf{\textsc{CheckUniqueness(sols, \(O\)})}
            \State \textbf{return} \textit{precisely exact-constrained}
        \Else
            \State \textbf{return} \textit{precisely underconstrained}
        \EndIf
    \EndProcedure
    \end{algorithmic}
\end{algorithm}
\noindent \textbf{\texorpdfstring{$K$}{4}-coefficient linear.}
Some circuits are inherently nonlinear but can be treated as linear and represented in matrix form to improve verification efficiency. These equations are considered linear because their higher-order terms contain one or more known variables. For such circuits, we employ the heuristic approach to identify special inputs that may cause underconstrained conditions (shown in Algorithm \ref{algo:check-multi-solutions}), thereby improving efficiency. Afterwards, a \texorpdfstring{$k$}{4}-coefficient linear circuit can be succinctly represented by a coefficient matrix \(A\) and a constant vector \(b\), and checked as a linear case. The pseudo-code of checking $k$-coefficient circuits is given in Algorithm \ref{algo:k_coeff_check}.
\begin{algorithm}[!ht]
    \caption{Check multiple solutions by analyzing coefficients of parameters}
    \label{algo:check-multi-solutions}
    \begin{algorithmic}[1]
        \Procedure{GetUndeterminedCoeffSols}{$K, U, E$}
            \State \(UCSolCountTable \gets EmptyHashTable\)
            \ForAll{$eq \in E$} \Comment{iterate all the equations}
                \State \(sols \gets \textsc{SolveUndeterminedCoefficientEq}(eq, K, U)\)
                \If{$sols \neq \emptyset$} \Comment{If there exists solution}
                    \ForAll{$sol \in sols$}
                        \State \(UCSolCountTable[sol] \gets UCSolCountTable[sol] + 1\)
                    \EndFor
                \EndIf
            \EndFor
            \State \textbf{return} \(UCSolCountTable\)
        \EndProcedure
    \end{algorithmic}
\end{algorithm}
\begin{algorithm}[!ht]
    \caption{Algorithm for checking $k$-coefficient circuits} \label{algo:k_coeff_check}
    \begin{algorithmic}[1]
        \Procedure{\textsc{CheckKCoeffcientCircuit}}{$\langle C(\mathcal{X}), E \rangle$}
        \State \(udcSols \gets \textsc{GetUndeterminedCoeffSols}(K, U, E)\)
        \ForAll {$sol \in udcSols$}
            \State \(E_{sub} \gets \textsc{SubstitueEqs(E, sol)}\)
            \If{Check \(C(U, K, O, E_{sub})\) is \textit{underconstrained}}
                \State \textbf{return} \textit{precisely underconstrained}
            \Else
                \State \textbf{continue}
            \EndIf
        \EndFor
        \State \(A, b \gets \textsc{LinearEquationsToMatrix}(E, U)\) 
        \State \(sols \gets \textsc{GaussJordanSolve(A, b)}\)
        \If{\textsc{IsEmpty}(sols)}
            \State \textbf{return} \textit{algebraically overconstrained}
        \ElsIf{\textsc{CheckUniqueness(sols, \(O\))}}
            \State \textbf{return} \textit{algebraically exact-constrained}
        \ElsIf{$\textsc{Size}(sols) > 1$}
            \State \textbf{return} \textit{precisely underconstrained}
        \Else
            \State \textbf{return} \textit{unknown}
        \EndIf
        \EndProcedure

    \end{algorithmic}
\end{algorithm}

\smallskip
\noindent \textbf{Higher-Order.}
For those higher-order circuits which cannot be treated as linear, the previous heuristic approach(Algorithm \ref{algo:check-multi-solutions}) will still be used to identify input variables that may cause the circuit to become underconstrained. The coefficients of the unknown variables are treated as equations and their solutions will be stored in a frequency table for counting the occurrences of each solution. Then, the solver sorts the solutions based on their frequencies and iteratively substitutes them into the original system of equations to obtain a new system, which is then solved to check if the circuit is underconstrained. If the input variable causing the circuit to become underconstrained cannot be identified, the system of equations will be directly solved with Gr\"{o}bner basis to verify the constrained situation of the circuit. The pseudo-code of checking polynomial circuits is given in Algorithm \ref{algo:higher_check}.
\begin{algorithm}[!ht]
    \caption{Algorithm for checking higher-order circuits} \label{algo:higher_check}
    \begin{algorithmic}[1]
        \Procedure{\textsc{CheckHigherCircuit}}{$\langle C(\mathcal{X}), E \rangle$}
            \State \(udcSols \gets \textsc{GetUndeterminedCoeffSols}(K, U, E)\)
        \ForAll {$sol \in udcSols$}
            \State \(E_{sub} \gets \textsc{SubstitueEqs(E, sol)}\)
            \If{Check \(C(U, K, O, E_{sub})\) is \textit{underconstrained}}
                \State \textbf{return} \textit{precisely underconstrained}
            \Else
                \State \textbf{continue}
            \EndIf
        \EndFor
        \State \(sols \gets \textsc{SolveHigherSystemWith$F_5$}(C)\)
        \If{\textsc{IsEmpty}(sols)}
            \State \textbf{return} \textit{algebraically overconstrained}
        \ElsIf{\textsc{CheckUniqueness(sols, \(O\))}}
            \State \textbf{return} \textit{algebraically exact-constrained}
        \ElsIf{$\textsc{Size}(sols) > 1$}
            \State \textbf{return} \textit{precisely underconstrained}
        \Else
            \State \textbf{return} \textit{unknown}
        \EndIf
        \EndProcedure
    \end{algorithmic}
\end{algorithm}
\subsection{Complexity Analysis}

This section examines the time complexity associated with each category of circuits. For linear circuits, including those characterized by \texorpdfstring{$K$}{4}-coefficient linear circuits, the Gauss-Jordan elimination method has a time complexity of \(O(n^3)\)~\cite{ANDREN2007428}. Our statistical analysis of Circomlib~\cite{Iden3CircomlibLibrary} indicates that the majority of cases involve linear circuits. Consequently, we anticipate that our algebraic-based method will significantly enhance performance compared to SMT solvers. In the case of higher-order circuits, we employed the \(F_5\) algorithm, which is designed for computing Gr\"{o}bner bases.\cite{f5algorithm} This algorithm typically exhibits $O(2^n)$ or $O(2^{2^n})$ time complexity in the worst-case scenarios. However, the \(F_5\) algorithm demonstrates substantial improvements over its predecessors, such as Buchberger’s algorithm\cite{BUCHBERGER2006475} and the \(F_4\) algorithm.

\section{Evaluations}\label{Eval}

\subsection{Environment Setting and Metrics}

\noindent\textbf{Environment configuration.}
The evaluations were performed on a machine with a 12th Gen Intel(R) Core(TM) i5-12500H processor and 16GB of DRAM.

\smallskip
\noindent\textbf{Check Benchmark.}
We settle a representative benchmark to evaluate the effectiveness and accuracy of \(\text{AC}^4\) ,Picus~\cite{10.1145/3591282} and CIVER~\cite{CIVER}. The benchmark was based on the data set collected from Circomlib, the standard library from Circom. All cases of our benchmark were divided into two parts: \textbf{Circomlib-utils benchmarks}, which was extracted from the Circomlib utils library and had a smaller size, and \textbf{Circomlib-core benchmarks}, which was extracted from the Circomlib core library and had a larger size~\cite{10.1145/3591282}.

Furthermore, we established a benchmark to evaluate the effectiveness and accuracy of AC\textsuperscript{4} and the halo2-analyzer\cite{halo2analyzer}, as implemented in the study by Soureshjani et al. (2023) \cite{SoureshjaniHJKG23}. This benchmark was constructed using a dataset of classical halo2 circuits, which were collected from various GitHub projects.

The primary benchmark focuses on circom datasets, while our results also demonstrate improvements on halo2 datasets compared to other tools.

\noindent\textbf{Benchmark Metrics}
\label{CR}
\noindent\textbf{solved rate.} The solved benchmarks are initially categorized into three types: precisely solved, algebraically solved, and unknown. The precisely solved cases are further categorized into three types: \textit{precisely exact-constrained}, which means that the circuit is safe; \textit{precisely overconstrained}, indicating that the circuit is overconstrained; and \textit{precisely underconstrained}, which suggests that the circuit is underconstrained. We omit this metric as $PS$ (Precisely Solved).

On the other hand, we propose a more relaxed metric, $AS$ (Algebraic Solved), by remapping cases that failed to obtain precise results into two new categories: \textit{algebraic exact-constrained} and \textit{algebraic overconstrained}, to reject \textit{overconstrained} and \textit{underconstrained} respectively. Note that \(\text{AC}^4\) still returns \textit{unknown} for extremely computationally exhausting cases.

\smallskip
\noindent\textbf{Checking time.} In this metric, our emphasis lies on the benchmark data on the parts that both \(\text{AC}^4\) and other tools can solve, and then proceed to compare the checking time of the tools.

\subsection{Detailed Evaluation over Circom Circuits}
\noindent\textbf{The Conditions of the Data Set over Circom}

\medskip\noindent\textbf{The size condition.}
Table \ref{tab:size_cond} presents the size condition of the benchmark set. Notably, within the benchmark set \textbf{Core}, there are three circuits with over 200,000 constraints, which are believed to be unsolvable using standard computers. Interestingly, in these three cases, only \textbf{Sha256compression} results in an out-of-memory exception due to the presence of around 20,000 unknown variables, resulting in the creation of a very large matrix. In contrast, the other two cases each have only one known variable, making the circuits easier to check.

\begin{table*}[!ht]
\centering
\caption{The size condition of the Circom benchmark sets}
\label{tab:size_cond}
\resizebox{\columnwidth}{!}{
\begin{tabular}{lrrrr}
\hline
 Benchmark set   &   \# of circuits &   Avg. \# of expr\_num &   Avg. \# of unknown\_num &   Avg. \# of out\_num \\
\hline
 utils           &              66 &                 1992 &                      12 &                   9 \\
 core            &             110 &                 6941 &                     200 &                  30 \\
 total           &             176 &                 5085 &                     130 &                  22 \\
\hline
\end{tabular}
 }
\end{table*}

We categorize circuits based on their constraint numbers in Table \ref{tab:size_cate_cond}, considering those with a constraint number less than 100 as ``small'' circuits, those with a constraint number in the range \([100, 1000)\) as ``medium'' circuits, and those with a constraint number in the range \([1000, \infty)\) as ``large'' circuits. This classification, which may be rudimentary, does not consider output signals. It is evident that small circuits can be solved directly, whereas other circuits may require specific restrictions to be solved within an acceptable time frame.
\begin{table}[!ht]
\centering
\caption{The size category condition of the circuits}
    \label{tab:size_cate_cond}
\begin{tabular}{lrrr}
\hline
 Benchmark set   &   utils &   core &   total \\
\hline
 small           &      49 &     63 &     112 \\
 medium          &       8 &     24 &      32 \\
 large           &       9 &     23 &      32 \\
\hline
\end{tabular}
\end{table}

\medskip
\noindent\textbf{The type condition.}
As shown in the table \ref{tab:ck_rate}, The benchmark set comprises 176 circuits, distributed as 71.6\% precisely linear, 6.2\% \texorpdfstring{$K$}{4}-coefficient linear, and 22.2\% quadratic, mirroring the range of circuit types encountered in real-world applications. Upon further examination of the \texorpdfstring{$K$}{4}-coefficient and quadratic cases, it becomes evident that a significant portion of them are binary circuits associated with cryptographic algorithms. Instead of exhaustive arithmetic solutions, focusing on binary optimization strategies offers promising efficiency gains in checking these circuits. For instance, the circuit \textbf{Num2BitsNeg@bitify\_256} is a quadratic circuit designed to convert an integer over a finite field into a 256 bit-vector, with all its unknown variables constrained by \(o \times (o -1 ) = 0\), indicating that they are strictly bits with their value limited to 0 or 1. Appropriately identifying these circuits and applying binary methods for solving them represents a more efficient approach compared to arithmetic methods.

\medskip
\noindent\textbf{Benchmark Results over Circom}
\smallskip
\noindent\textbf{Results based on Circuit Size.}
Table~\ref{tab:ck_time} presents the results across circuits of different sizes, highlighting the relationship between benchmark scale and solving difficulty. In general, larger benchmarks exhibit greater solving difficulty, leading to longer solution times and lower checking rates.However, in the \texttt{circomlib-utils} benchmark, the ``small'' category exhibits a higher average solving time. This is because two small circuits, \textbf{BitSub@binsub} and \textbf{BitSum@binsum}, which are quadratic, take unusually long to solve (128.26\,s and 185.88\,s, respectively). If we exclude these two circuits, the average PS time and the average PS\&AS time drop to 0.04\,s and 0.89\,s, respectively.

In some large circuits, despite a high number of constraints, the number of output signals is limited to 1, allowing for a relatively easy check once the file read operation concludes. Nonetheless, there exist extremely large circuits that are characterized by a linear type, yet their extensive constraint size renders the Gaussian-Jordan method challenging to execute, leading to out-of-memory issues. Conversely, smaller cases are relatively straightforward to solve directly but due to limitations, the precise solution is not attainable, resulting in algebraic outcomes.

Fortunately, the majority of practical Circom circuits are of the precisely linear type. This characteristic is advantageous because medium or large circuits that are not precisely linear cannot be solved directly due to unknown outcomes. Therefore, alternative methods must be utilized to bypass the direct solution of non-linear circuits. Additionally, providing a more accurate range to determine whether to solve the circuits can help avoid obtaining algebraically solved results. This approach is crucial for maintaining clarity and precision in circuit analysis.

\begin{table*}[!ht]
\caption{Results of $AC^4$ categorized by circuit size, including the checking time and solved rate for precisely solved and algebraically solved cases. The terms K, U, O, v, PS, AS represent known, unknown, output, variables, precisely solved, and algebraically solved, respectively. The explanation of PS, AS is discussed in Section \ref{CR}.}\label{tab:ck_time}
\centering
\resizebox{1\columnwidth}{!}{
\begin{tabular}{ccccccccccc}
\hline
\multicolumn{2}{c}{{Benchmark}}     & \multicolumn{4}{c}{{Utils}}      & \multicolumn{4}{c}{{Core}}     & \multirow{2}{*}{{overall}} \\ 
\cline{1-2} \cline{3-6} \cline{7-10}
\multicolumn{2}{c}{{Size}}                                 & {small} & {medium} & {large} & {overall} & {small} & {medium} & {large} & {overall} &                                   \\ \hline
\multirow{5}{*}{{Avg. \# variable}} & {K}           & 15             & 424             & 14309          & 2014             & 36             & 435             & 24984          & 5159             & 3973                              \\
                                           & {U}           & 4              & 18              & 58             & 13               & 11             & 78              & 26             & 29               & 23                                \\
                                           & {O}           & 2              & 2               & 57             & 9                & 10             & 78              & 26             & 29               & 21                                \\
                                           & {total v}     & 18             & 442             & 14366          & 2026             & 47             & 512             & 25010          & 5188             & 3995                              \\
                                           & {constraints} & 14             & 244             & 14230          & 1993             & 17             & 474             & 24923          & 5144             & 3956                              \\ \hline
\multicolumn{2}{c}{{Avg. PS Time(s)}}                      & 7.89           & 0.20            & 7.03           & 6.79             & 0.10          & 1.36             & 10.94          & 2.55             & 1.08                              \\
\multicolumn{2}{c}{{Avg. AS\&PS Time(s)}}                  & 7.27          & 1.69            & 6.40           & 6.47             & 0.09           & 1.36             & 10.06          & 2.38             & 1.72                             \\ \hline
\multicolumn{2}{c}{{ps\_rat}}                      & 0.73           & 0.88            & 0.89           & 0.77             & 0.83           & 1.00             & 0.91          & 0.88             & 0.84                              \\
\multicolumn{2}{c}{{ps\&as\_rat}}                  & 0.92           & 1.00            & 1.00           & 0.94             & 0.92           & 1.00             & 1.00          & 0.95             & 0.97                              \\ \hline

\end{tabular}
}
\end{table*}

\smallskip \noindent \textbf{Result based on Circuit Type.}
Table~\ref{tab:ck_rate} presents the results across circuits of different types, the data indicates a significantly high solved rate for precisely linear circuits. However, Substantial constraints in circuits result in very large augmented matrices. As a consequence, the checke rate of \(k-\)coefficient circuits is relatively low. This is attributed to the adoption of an algebraic strategy for \(k-\)coefficient circuits, which in turn leads to the direct solving of small-size \(k-\)coefficient circuits without returning a precise result.

Heuristic algorithms initially assess potential underconstraint in quadratic circuits. We then apply classic algorithms for solving quadratic equations over finite fields to obtain direct solutions. Moreover, in the case of specific binary circuits such as \textbf{BinSum} and \textbf{Num2Bits}, a detection method is utilized to verify the boolean nature of unknown variables. Upon confirmation, a specialized method for solving binary circuits is employed in lieu of arithmetic-based approaches. Most circuits are thoroughly verified before processing.

\begin{table*}[!ht]
\caption{Results of $AC^4$ categorized by circuit type, including the checking time and solved rate for precisely solved and algebraically solved cases. We categorize circuits based on their types: pl means precisely linear type, kcl means k coefficient linear, q means quadratic, ps means precisely solved, as means algebraically solved.} \label{tab:ck_rate}
\centering
\scalebox{.9}{
\begin{tabular}{cccccccccc}
\hline
{Benchmark}   & \multicolumn{4}{c}{{Utils}}                         & \multicolumn{4}{c}{{Core}}                          & \multirow{2}{*}{{overall}} \\ \cline{1-9}
{type}        & {pl} & {kcl} & {q} & {overall} & {pl} & {kcl} & {q} & {overall} &                                   \\ \hline
{ps\_num}     & 45          & 3            & 7          & 55               & 75          & 3            & 23         & 101              & 156                               \\
{as\_num}     & 3           & 3            & 5          & 11               & 2           & 2            & 4          & 8                & 19                                \\
{total\_num}  & 48          & 6            & 12         & 66               & 77          & 5            & 27         & 109              & 175                               \\ \hline
{ps\_rat}     & 94\%        & 60\%         & 58\%       & 83\%             & 96\%        & 60\%         & 85\%       & 91\%             & 89\%                              \\
{ps\&as\_rat} & 100\%       & 100\%        & 100\%      & 100\%            & 98\%        & 100\%        & 100\%      & 99\%             & 99\%                              \\ \hline
{Avg. PS Time(s)}     & 1.30        & 0.05         & 44.93       & 6.79             & 3.00        & 0.03         & 1.39       & 2.55             & 2.05                              \\
{Avg. AS\&PS Time(s)} & 1.50       & 0.43        & 29.37      & 6.47            & 2.96       & 0.04        & 1.19      & 2.38             & 2.16                              \\ \hline
\end{tabular}
}
\end{table*}
\medskip

\noindent\textbf{Comparison with Picus and CIVER}


We compared our tool, $AC^4$, with Picus and CIVER using the \texttt{circomlib-utils} and \texttt{circomlib-core} benchmarks.
For Picus, the average checking times for small, medium, and large circuits are 1.95 seconds, 6.32 seconds, and 3256.48 seconds, respectively, with corresponding solving rates of 80.24\%, 48.32\%, and 68.55\%. The overall average checking time and solving rate are 365.52 seconds and 69.36\%, respectively.
For CIVER, the average checking times for small, medium, and large circuits are 0.85 seconds, 2.58 seconds, and 6.79 seconds, respectively, while the corresponding solving rates are 61.32\%, 46.66\%, and 58.33\%. The overall average checking time and solving rate for CIVER are 1.19 seconds and 58.13\%, respectively.

\begin{figure}[!ht]
    \centering
    \subfloat[Comparison between the checking time of \(\text{AC}^4\), Picus and Civer. ]{
        \label{fig:ck_time}
        \includegraphics[width = .45\linewidth]{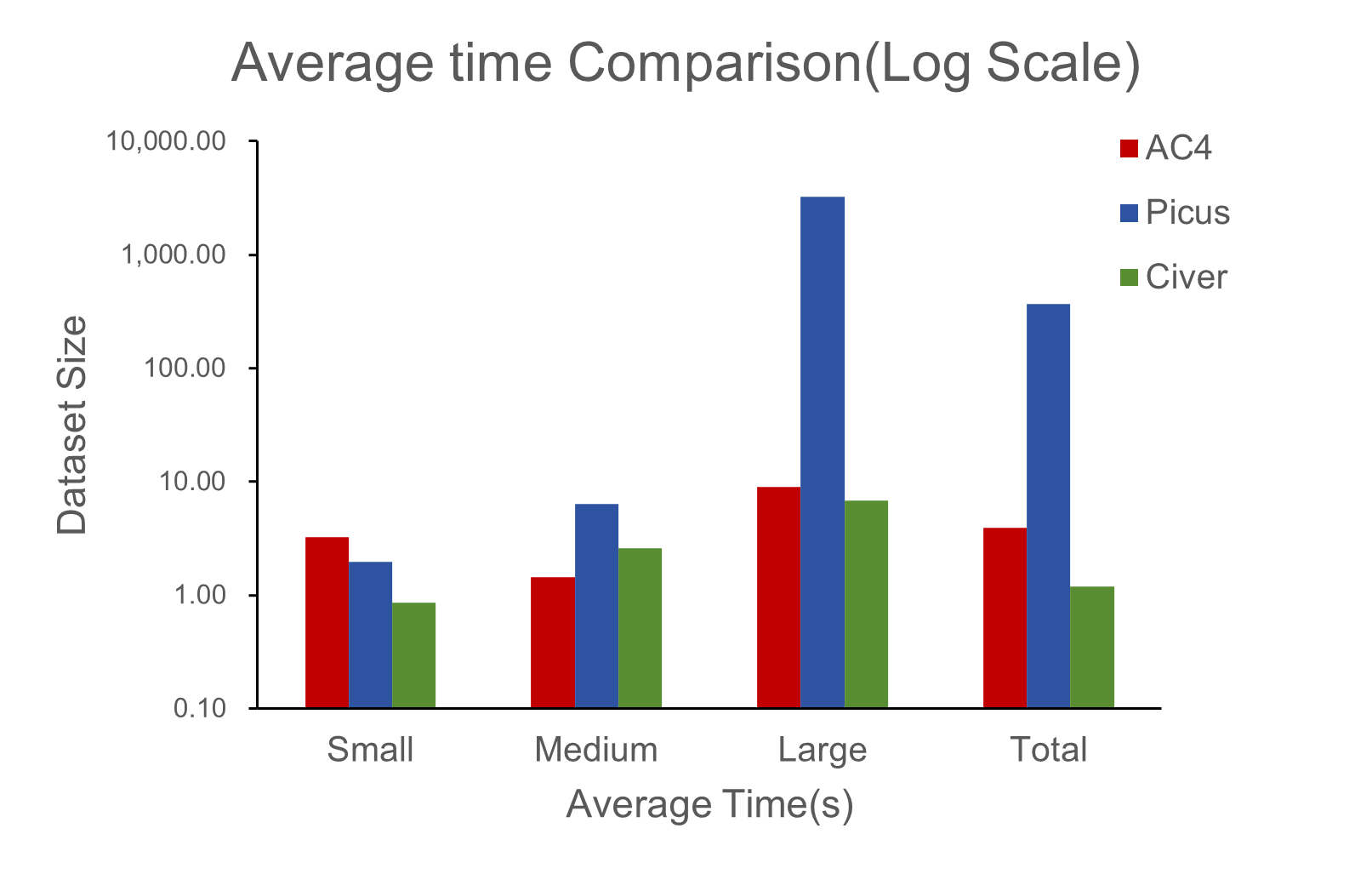}
    }\hspace{2em}
    \subfloat[Comparison between solved rate of \(\text{AC}^4\), Picus and Civer.]{
        \label{fig:ck_rate}
        \includegraphics[width = .45\linewidth]{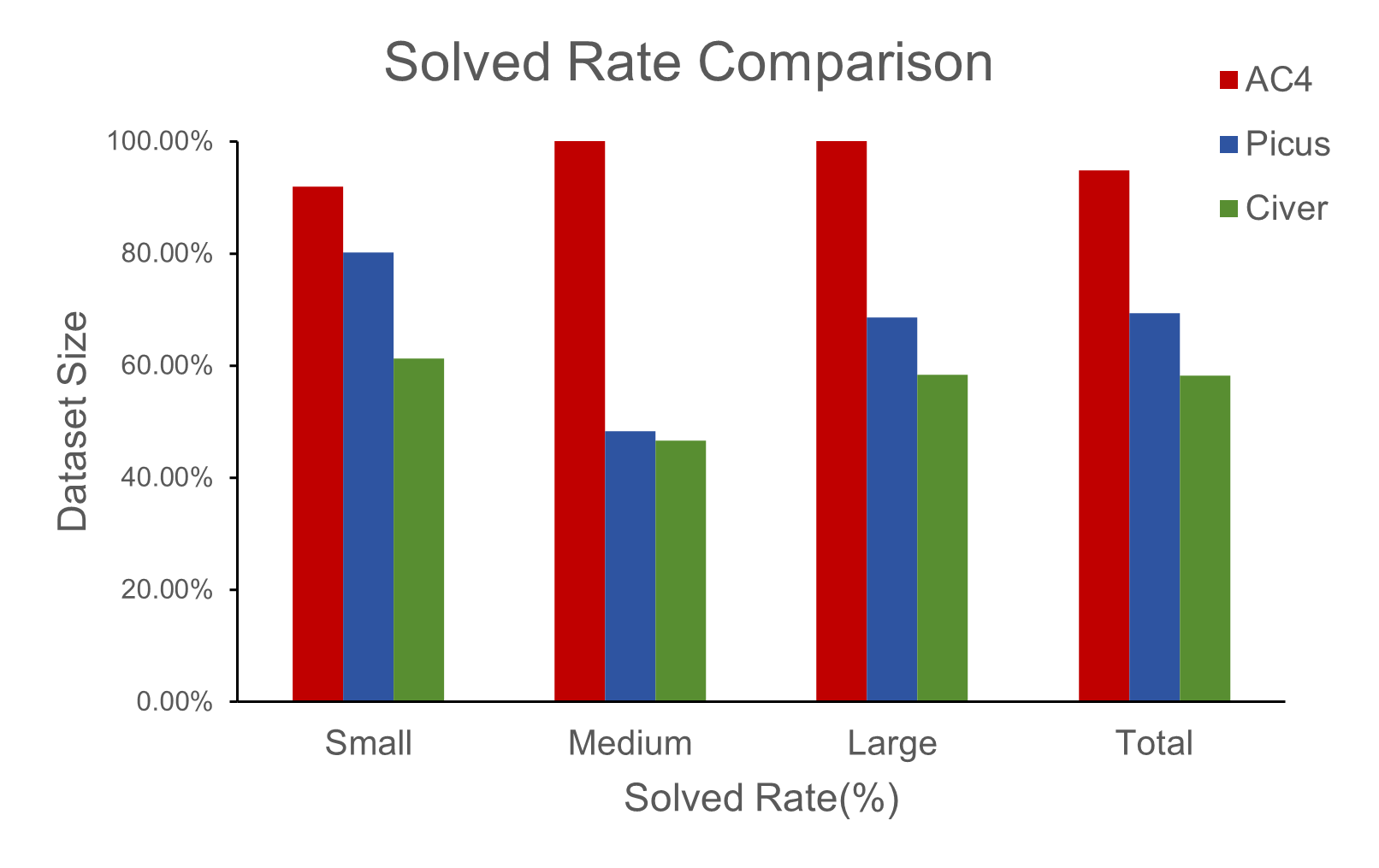}
    }
    \label{}
    \caption{The checking time and solved rate comparisons between \(\text{AC}^4\), Picus and Civer}
\end{figure}

As a comparison with $AC^4$, the checking time results are shown in Fig.~\ref{fig:ck_time}, and the checking rate results are shown in Fig.~\ref{fig:ck_rate}. The results demonstrate that our tool achieves the highest checking rate among the three tools, particularly on medium- and large-sized circuits, indicating its superior capability in solving complex circuits. However, $AC^4$ exhibits a relative disadvantage in solving time for small and large circuits compared with CIVER. This is primarily because CIVER imposes a strict timeout limit, which causes it to skip overly complex circuits. These timeouts occur more frequently in large size circuits, leading to shorter average solving times but a lower overall checking rate. For the small circuits, the higher average time is largely due to two circuits, \textbf{BitSub@binsub} and \textbf{BitSum@binsum}. If we exclude these two circuits, the average time drops to 0.42\,s, which is faster than CIVER.

\subsection{Evaluation over Halo2 Circuits}
In the absence of a standard library akin to Circomlib for Halo2, we meticulously selected 29 representative circuit implementations from several prominent open-source GitHub projects. These implementations encompass essential application scenarios, including large number arithmetic, multiplication and inner product, the Poseidon hash algorithm, and Ethereum Virtual Machine operations. The dataset is specifically categorized as follows: 10 test cases for small-scale circuits, 11 test cases for medium-scale circuits, and 8 test cases for large-scale circuits.

This curated dataset was then utilized to construct a comprehensive dataset for analysis. Using this dataset, we performed benchmark analyses in conjunction with the halo2-analyzer.

\begin{figure}[!ht]
    \centering
    \subfloat[Comparison between the checking time of \(\text{AC}^4\) and halo2-analyzer. ]{
        \label{fig:halo2_ck_time}
        \includegraphics[width = .45\linewidth]{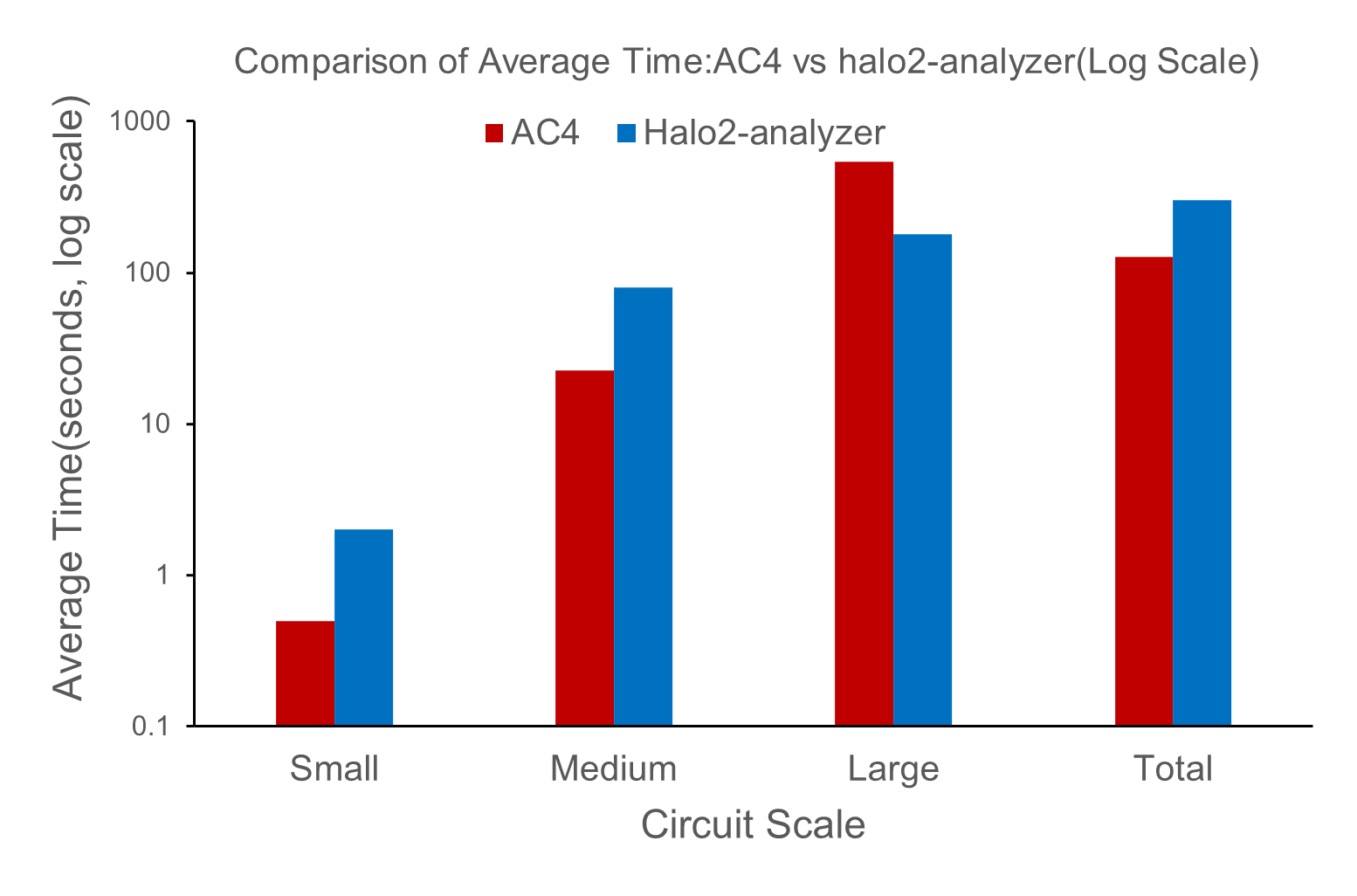}
    }\hspace{2em}
    \subfloat[Comparison between solved rate of \(\text{AC}^4\) and Halo2-analyzer]{
        \label{fig:halo2_rate}
        \includegraphics[width = .45\linewidth]{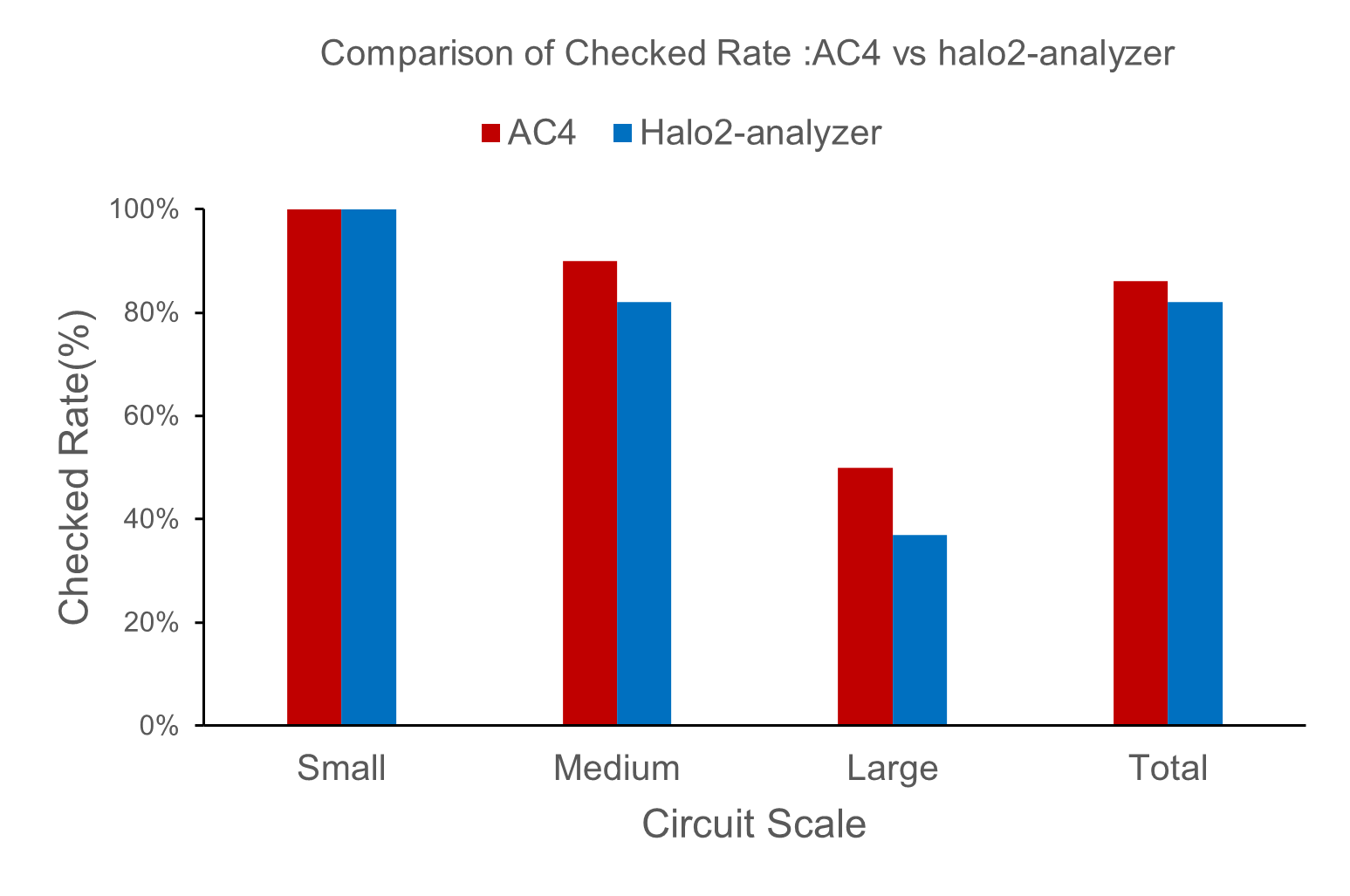}
    }
    \caption{The checking time and solved rate comparisons between \(\text{AC}^4\) and halo2-analyzer}
    \label{fig:halo2_time_rate}
\end{figure}

As shown in figure \ref{fig:halo2_time_rate}, AC\texorpdfstring{$^4$}{4} demonstrates a enhancement in efficiency and accuracy compared to halo2-analyzer, exhibiting an improvement over efficiency and accuracy.

\subsection{Evaluation Result Analysis}
Table \ref{tab:ck_time} illustrates a significant correlation between the size of the circuit and the time required for checking and solved rate. Despite their size, large circuits keep checking times low due to two critical factors. Firstly, the predominance of linear circuits, efficiently addressed by the Gauss-Jordan solver, contributes to the efficiency in solving a majority of cases. Additionally, optimizations targeting binary cases within k-coefficient and higher-order circuits serve to further reduce the solving time despite their slightly larger size.

The solved rate is high for precisely linear cases; however, there are some cases for which we cannot provide precisely checked results due to the lack of an algebraic solution. Special input scenarios can reduce the augmented matrix's rank, yielding a solution. In contrast, for the \texorpdfstring{$K$}{4}-coefficient cases, the solved rate is relatively low, even when a mixed solver is employed. This is because we did not implement certain static analysis methods to elaborate the benchmark set, which would have facilitated the checking of circuits. Conversely, the solved rate is relatively high for quadratic cases, as most of these cases are in binary form. The detection method and optimization for binary quadratic circuits have significantly enhanced efficiency in checking these cases.

A more intriguing observation is that Table \ref{tab:ck_rate} checking time for the precisely linear type is unexpectedly large. This occurs because the distribution of circuit types in the benchmark is highly unbalanced—over 71.6\% of the circuits are classified as precisely linear. Consequently, this category contains a larger proportion of large-sized circuits, which leads to the observed increase in average checking time. 

In the context of quadratic cases, Picus and halo2-analyzer employ the cvc5-ff-range solver. This solver first implements finite field theory and then utilizes Buchberger's algorithm along with triangular decomposition to simplify polynomial equations. To address the problem, it further applies Collins' cylindrical algebraic decomposition (CAD)~\cite{Collins-CAD} \cite{cvc5-ff}. It is important to note that the time complexity of CAD is \(O(2^{2^n})\), which can be prohibitive in many scenarios involving large quadratic cases. Additionally, the \(F_5\) algorithm demonstrates superior performance over modular integers in various cryptographic challenges.

\smallskip \noindent \textbf{AS Analysis.} 
We point out that the cases that occur in the AS metric are worth discussing. In \textbf{Core} data split, cases that cannot be precisely solved consist of nearly 7\% of the total amount, while this number increases to 17\% in \textbf{Utils} data split. Our proposed AS mechanism enables our tool to investigate more detailed situations of constraints. By excluding an incorrect option among underconstrained and overconstrained, we narrow the possible status of these cases. Experimental results demonstrate that our tool offers valuable hints to almost 100\% cases in both  \textbf{Core} and \textbf{Utils} splits.

\smallskip \noindent \textbf{Limitations.} 
Picus utilized their algorithm \textit{Uniqueness Constraint Propagation} to solve all the \(k-\)coefficient circuits. The algorithm simplifies verification/resolution by strategically assigning values to unknown variables through partial equation analysis, thereby reducing the number of unknowns. In contrast, our method forgoes static analysis, which limits the checking of the \texorpdfstring{$K$}{4}-coefficient cases.

\smallskip \noindent \textbf{Analysis over halo2 circuits.}
Halo2-analyzer employs cvc5-ff to verify circuit uniqueness, giving our tool an advantage in solving large-scale polynomial equation systems.

\section{Related Work}

\noindent \textbf{Verification on ZK programs.} 
Circomspect~\cite{Circomspect} serves as a static analyzer and linter designed for the Circom programming language. Picus~\cite{10.1145/3591282} employs a static analysis method combined with an SMT solver. Circomspect and Picus share certain similarities with our work. However, Circomspect can only annotate a portion of underconstrained variables.

ZK circuit formalization has seen notable advancements: Shi et al.~\cite{shi2023dataflowbased} proposed a data-flow-based algorithm for R1CS standardization, to enhance performance across R1CS tasks. Wen et al.~\cite{cryptoeprint:2023/190} formalized and classified Circom vulnerabilities, introducing a CDG-based static analysis framework. CODA~\cite{cryptoeprint:2023/547} emerged as a statically typed language for formal ZK system property specification and verification. Soureshjani et al.\cite{SoureshjaniHJKG23} applied abstract interpretation to examine halo2 circuit properties, on the Plonkish matrix using an SMT solver. 

Research on ZK compilers with theorem proving is ongoing. Bangerter and Almeida et al.~\cite{cryptoeprint:2008/471,10.1007/978-3-642-16441-5_5,10.1007/978-3-642-15497-3_10,bangerterdesign}, pioneered verifiable compilation using Isabelle/HOL~\cite{ref1}. PinocchioQ~\cite{7536381} verified Pinocchio\cite{10.1145/2856449}, a C-based zk-SNARK, with Compcert~\cite{10.1145/1111037.1111042} based on Coq~\cite{Bertot2004}. Chin and Coglio et al~\cite{cryptoeprint:2023/1278,Coglio_2023}. employed ACL2~\cite{588534} on Leo, while Bailey et al.~\cite{cryptoeprint:2023/656} established formal proofs for six zk-SNARK constructions using the Lean theorem prover~\cite{10.1007/978-3-319-21401-6_26}.

\smallskip
\noindent\textbf{Algebra computation in circuit verification.} Algebraic computation is a long-standing but niche endeavor in verification. Watanabe~\cite{4601876} utilized Gr\"{o}bner Basis~\cite{FAUGERE199961,f5algorithm} and polynomial reduction techniques to verify arithmetic circuits in computing and signal processing systems. Similarly, Lv et al.~\cite{6113989,6167783} employed a computer-algebra-based approach for the formal verification of hardware. Scholl et al.~\cite{9218721} advanced this field by combining Computer Algebra with a Boolean SAT solver to verify divider circuits. In the context of Zero Knowledge Proof (ZKP) compilers, Alex et al.~\cite{DBLP:conf/cav/boundedFFB} presented a partial verification of a field-blasting compiler pass. Furthermore, Thomas et al.~\cite{DBLP:conf/lpar/HaderRK23} introduced a novel automated reasoning method for determining the satisfiability of non-linear equation systems over finite fields. Additionally, Thomas et al.~\cite{DBLP:conf/ijcar/HaderKIGK24} enhanced the Yices2 SMT solver by implementing a new MCSat-based reasoning engine for finite fields.

Alex et al.~\cite{cvc5-ff} employed the C++ CLN library \cite{cln-cpp-lib} to implement finite fields and utilized COCOALib's \cite{COCOA-lib} implementation of the Buchberger algorithm for preprocessing equations, subsequently applying CAD algorithm\cite{Collins-CAD} to solve them and presents a new SMT solver for finite field equations that improves efficiency on cryptosystem verification tasks by using multiple simplified Gr\"{o}bner bases and specialized propagation algorithms\cite{DBLP:conf/cav/SplitGrobnerBases}.

We emphasize that much of the work in this field reduces program or hardware issues to algebra. To our knowledge, we are the first to apply computer algebra methods in ZKPs.

\section{Conclusion}

Using an algebraic computation system, we have developed a technique for detecting zero-knowledge bugs in arithmetic circuits. This approach is designed to address issues caused by underconstrained or overconstrained arithmetic circuits. Unlike the traditional method of encoding the problem into formulas and utilizing an SMT solver, our approach leverages arithmetic computations, significantly improving efficiency. Additionally, to enhance the verification process, we have incorporated heuristic methods. Notably, our method directly analyzes arithmetic circuits and does not depend on a specific domain-specific language (DSL), making it applicable to a wide range of DSLs that support zk-SNARKs. Our implementation utilizes the open-source algebra computation system, \textit{SymPy}, in a tool called \(\text{AC}^4\). We conducted an evaluation involving $176$ Circom circuits and $29$ halo2 circuits, and our approach successfully verified approximately 89\% of the benchmark sets within a short timeframe.

In order to enhance the elaboration of the data set prior to direct solution, our approach will encompass adopting methods such as static program analysis. This will enable us to acquire comprehensive information about the circuits before applying our methods. Additionally, we aim to conduct extensive evaluations of commonly utilized circuits and design specialized algorithms for them. Furthermore, the use of theorem proving will be employed to demonstrate the correctness of a series of specific circuits. Ultimately, our goal is to develop or modify existing open-source computer algebra tools to tailor a solver for determining the number of solutions to a system of polynomial equations over a finite field.

\bibliographystyle{splncs04}
\bibliography{ac4}

\end{document}